%% file: lexicalratio.tex
\documentclass[12pt,letterpaper]{article}
\usepackage{geometry}
\usepackage[cp1252]{inputenc}
\usepackage[T1]{fontenc}
\usepackage[scale=0.96]{XCharter} 
\usepackage{lmodern}
\usepackage{graphicx}
\usepackage{amsmath}
\usepackage{amssymb}
\usepackage{amsthm}
\usepackage{amsfonts}
\usepackage{authblk}
\usepackage{epstopdf}	
\usepackage{verbatim}
\usepackage{multicol}
\usepackage{tabularx}
\usepackage{natbib}
\usepackage{graphicx}
\usepackage{multirow}
\usepackage{array}
\usepackage{color}
\usepackage{float}
\restylefloat{table}
\usepackage{xcolor}
\usepackage{lscape}
\usepackage{ragged2e}
\usepackage{booktabs}
\usepackage{topcapt}
\usepackage{rotating}
\usepackage{setspace}
\usepackage{txfonts}
\usepackage{bm}
\usepackage{titlesec}
\usepackage{indentfirst}
\usepackage{enumitem}
\usepackage[pdftex,breaklinks,colorlinks,citecolor=black,urlcolor=black,linkcolor=black]{hyperref}
\usepackage{tikz}
\usetikzlibrary{arrows,positioning} 
\usepackage{algorithmic,algorithm}
\usepackage{dcolumn}
\usepackage{pifont}
\usepackage{authblk}
\usepackage{xr}
\usepackage{actuarialangle}
\usepackage{longtable}
\usepackage{xurl}

\definecolor{darkblue}{rgb}{0.0, 0.0, 0.55}
\hypersetup{colorlinks=true,linkcolor=darkblue,filecolor=darkblue,urlcolor=darkblue,citecolor=darkblue}

\input{info}

\definecolor{alizarin}{rgb}{0.82, 0.1, 0.26}

\newcolumntype{Y}{>{\raggedleft\arraybackslash}X}
\newcolumntype{C}{>{\centering\arraybackslash}X}
\newcolumntype{d}[1]{D{.}{.}{#1}}

\allowdisplaybreaks
\date{}
\author{\Authors}
\title{\Title}

\titleformat{\section}{\normalfont\bfseries}{\thesection}{1em}{}
\titleformat{\subsection}{\normalfont\bfseries}{\thesubsection}{1em}{}
\titleformat{\subsubsection}{\normalfont\bfseries}{\thesubsubsection}{1em}{}

\titlespacing*{\section}{0pt}{6pt}{6pt}
\titlespacing*{\subsection}{0pt}{6pt}{6pt}
\titlespacing*{\subsubsection}{0pt}{6pt}{6pt}

\begin{document}
\makeatletter
\g@addto@macro{\normalsize}{%
    \setlength{\abovedisplayskip}{4pt}
    \setlength{\abovedisplayshortskip}{4pt}
    \setlength{\belowdisplayskip}{4pt}
    \setlength{\belowdisplayshortskip}{4pt}}
\makeatother

\maketitle
\thispagestyle{empty}
\vspace{-0.85cm}

\begin{abstract}
\Abstract \\ 


\noindent {\it JEL Classification:} \JEL
\end{abstract}

\newpage 
\setcounter{page}{1}
\doublespacing

\section{Introduction and motivation\label{sec:introduction}}

Portfolio diversification is a cornerstone of modern finance, designed to mitigate risk by spreading investments across a broad spectrum of financial instruments, asset classes, industries, and geographic regions. The underlying principle is that diversification can reduce the impact of any individual asset's poor performance on the overall portfolio, thereby enhancing its risk-adjusted returns. By investing in assets that are not perfectly related to one another, portfolio diversification helps to balance the trade-off between risk and return, providing a more stable performance over time.

The main metric used to measure diversification in finance is correlation, quantifying the degree to which two assets move in relation to each other. When assets have low or negative correlations, they tend to move independently or in opposite directions, which benefits diversification by reducing overall portfolio volatility. A portfolio with lower correlations is generally more diversified because the risk of large losses is reduced when some assets perform poorly while others perform well. One of the earliest and most influential works that formalized this concept is by \citet{markowitz1952}, who introduced mean-variance optimization. In his framework, the correlation between assets is critical because it determines how effectively risk can be minimized for a given level of expected return. By strategically combining assets with varying degrees of correlation, Markowitz demonstrated how investors could construct a portfolio that achieves the optimal balance between risk and return, resulting in the efficient frontier of portfolios.

The literature on portfolio diversification is rich. However, much remains to be done in the field. \citet{10.1093/rfs/hhm075} discussed the lack of robustness among existing diversification metrics and their poor performance on real-world out-of-sample data. This can be explained by the reliance on quantitative financial time series data as the sole source of information. In the recent literature, one of the most notable examples of a diversification index based on a general risk measure is the diversification ratio (DR), introduced by \citet{choueifaty2008toward}. The DR measure is a crucial metric that quantifies the effectiveness of diversification by measuring the ratio of the weighted average volatility of individual assets to the overall portfolio volatility. Studies by \citet{mainik2010}, \cite{degen2010risk}, and \citet{embrechts2009multivariate} have explored diversification metrics constructed using risk measures, including risk-based forms of DR. These studies particularly underscore the value of diversification in mitigating portfolio risk, even under extreme market conditions, where correlations tend to increase and diversification benefits can diminish.

Current metrics, constructed using risk measures or correlations, fail to capture the complexities of financial markets. The increasing availability of textual data related to financial assets, such as news articles, social media posts, and company reports, presents an opportunity to capture dimensions of diversification that may not be evident in numerical data alone. Entropy-based measures have emerged as a promising approach to quantify and enhance portfolio diversification. These measures, rooted in information theory, provide a framework for assessing the distribution of weights across assets in a portfolio. Shannon entropy, in particular, has been widely applied to portfolio selection, with higher entropy values indicating greater diversification \citep{kirchner2011measuring}. Our proposed metric is rooted in entropy concepts and will inherit the said desirable properties.

The incorporation of entropy in portfolio models has shown several advantages. \citet{ormos2013entropy}, among others, have demonstrated that entropy-based approaches can outperform traditional methods like the mean-variance model regarding Sharpe ratios and mean portfolio returns. Furthermore, entropy maximization has increased portfolio heterogeneity and made asset allocation more practicable \citep{bera2008portfolio}. However, entropy-based measures have many limitations. These measures act more as an optimization strategy to balance weights rather than a measure of diversification. In other words, they do not provide a mathematical function to quantify the diversification of a portfolio. Despite these drawbacks, entropy-based measures continue to attract attention in portfolio theory. Their ability to capture nonlinear dependencies and provide a more comprehensive view of portfolio risk makes them a valuable tool in the ongoing quest for improved diversification strategies. We will be leveraging many of these desirable properties in our proposed metric.

The application of textual analysis in finance has gained significant traction in recent years. The use of natural language processing (NLP) techniques and textual data has focused on a few common themes in finance, and to our knowledge, our proposed metric is the first linguistic metric of portfolio diversification. Work in the NLP field often focuses on sentiment analysis and asset return prediction. For example, \citet{kumar2016machine} surveyed machine learning applications in financial market prediction, including natural language processing for sentiment analysis of news articles and social media posts. Their work highlighted the potential of incorporating textual data into financial decision-making processes. More recently, \citet{shulman2020nlp} discussed the increasing adoption of NLP techniques in finance for tasks such as automating the capture of earnings calls, enriching data with context, and improving search and discovery across proprietary financial datasets. By leveraging NLP techniques, the lexical ratio (LR) measure aims to provide a unique perspective on portfolio composition and potential hidden correlations between assets that existing methods would not capture.

Overall, this paper offers three main contributions to the field. First, it introduces LR and describes its theoretical properties. Second, it identifies the drawbacks of relying on historical asset losses as the sole source of information and offers a new framework to rethink diversification measures. Third, it provides a comprehensive comparison of the performance of various metrics on real-world data and demonstrates the advantages of LR.

Given the unconventional nature of the LR measure, we start by describing the new measure in Section~\ref{sec:LR}. Section~\ref{sec:comparison} investigates the new measure and compares it with other approaches used to assess diversification. We show that there is significant information overlap between them, and we examine the robustness of various metrics and compare them with LR, demonstrating the limits of existing measures. We also conduct rigorous real-world data optimization tests on portfolios constructed from the S\&P 500 stocks. A discussion on possible extensions is provided in Section~\ref{sec:discussion}. Section~\ref{sec:conclusion} concludes.

\section{The lexical ratio\label{sec:LR}}

\subsection{Definition and calculation}

LR reimagines portfolio diversification by treating each asset as a document composed of words, such as news headlines or key financial terms. By analyzing the richness and balance of these words across all assets using Shannon's entropy, LR provides a unique perspective on how diversified a portfolio truly is. This approach captures the distribution of information in a way that traditional metrics may miss, especially when assets are influenced differently by external factors like news events.\
The key concept of the LR metric is that asset-related texts, such as news headlines or statements, contain valuable information about the asset's risks and potential benefits. The distribution of the words in such documents can be an invaluable source for examining these matters and combining various assets with different word distributions results in a new word distribution, which can provide insights into how the assets in a portfolio relate to one another. For example, having a diverse range of words from various sectors would be beneficial, while having words from a single sector indicates low diversification. News about specific political events does not affect all assets equally. Therefore, having a balanced set of words regarding different news, such as oil shipments or interest rates, could be invaluable.

LR is a portfolio's normalized Shannon entropy of combined weighted asset documents. It is calculated as 
\begin{equation}
\text{LR} = - \frac{1}{\log(m)} \sum_{k=1}^{m} \left(\frac{\sum_{i=1}^n w_i \, c_{i,k}}{\sum_{i=1}^n \sum_{j=1}^m w_i \, c_{i,j}}\right) \log \left(\frac{\sum_{i=1}^n w_i \, c_{i,k}}{\sum_{i=1}^n \sum_{j=1}^m w_i \, c_{i,j}}\right) ,
\end{equation}
where $m$ represents the number of distinct terms considered across all asset documents used for normalization, $n$ is the number of assets in a given portfolio, $w_i$ denotes the weight assigned to the $i^{\text{th}}$ asset in the portfolio, and $c_{i,k}$ is the number of occurrences of the $k^{\text{th}}$ term in the document associated with the $i^{\text{th}}$ asset.

This formulation offers intuitive insights into portfolio diversification by examining the distribution of words across asset-specific documents within the portfolio. The key idea is that just as a richer vocabulary and balanced word usage in the text indicate a well-rounded and diverse document collection, a portfolio with a diverse and balanced distribution of information (words) across its assets reflects a higher degree of diversification. The LR captures how varied the vocabulary is, representing the diversity of information sources or asset characteristics and how evenly the information is spread across assets. A higher LR suggests that no single asset dominates the portfolio's composition, leading to a more diversified and balanced investment. A lower LR may indicate a concentration in a few assets, reducing diversification. This measure offers a novel way to interpret how asset-specific information contributes to overall portfolio balance.

\subsection{Properties of LR}

LR's properties are essential to understanding its effectiveness as a portfolio diversification metric. These characteristics ensure that LR captures the complexities of asset distributions mathematically rigorously, building on the theoretical framework introduced earlier. By examining properties such as maximality, concavity, and scale invariance, it becomes clear how LR operates reliably across various portfolio configurations. These properties guarantee the robustness of LR and highlight why it provides a more nuanced view of diversification compared to traditional metrics. Understanding these aspects is critical for demonstrating the practical value of LR in real-world portfolio management.

Given that the LR measure is a normalized and weighted form of Shannon's entropy, it shares several key properties with Shannon's entropy, making it suitable for assessing portfolio diversification. We consider eight different properties related to Shannon's entropy and also shared by LR.
\begin{enumerate}
\setlength\itemsep{0em}
\item \textbf{Maximality}: LR reaches its maximum value when all distinct terms are equally represented in the combined asset documents, aligning with the intuitive notion of maximum diversification \citep{bera2008optimal}. This means that an optimal portfolio has a balanced distribution of words. For example, a balance between assets is affected by various news, such as interest rates, market sentiments, sectors, etc.
\item \textbf{Additivity}: For independent asset documents, the joint entropy is the sum of individual entropies, which can be useful when considering portfolios of independent asset classes \citep{gray2011entropy}. This allows LR to scale appropriately when combining independent sub-portfolios. This ensures consistency in diversification assessment and simplifies the analysis of complex portfolios composed of various asset classes.
\item \textbf{Concavity}: LR is a concave function of the term distribution, ensuring that diversification increases as terms become more evenly distributed \citep{cover2012elements}. This property rewards asset documents that offer diverse word collections.
\item \textbf{Continuity}: LR is a continuous function of the term distribution, ensuring smooth changes in the diversification measure as portfolio weights are adjusted \citep{gray2011entropy}. This provides stability, as small changes in asset weights lead to gradual and predictable shifts in the LR value.
\item \textbf{Non-negativity}: LR is always non-negative, ensuring easy interpretation of numbers \citep{cover2012elements}. A zero value indicates no diversification, while positive values indicate varying degrees of diversification, facilitating straightforward comparisons across portfolios.
\item \textbf{Symmetry}: LR is invariant under the permutation of the terms, meaning that the order of terms in the combined asset documents does not affect the diversification measure \citep{cover2012elements}. This ensures objectivity, as the measure remains consistent regardless of how information is ordered. This property is crucial for fair and unbiased diversification assessment.
\item \textbf{Expansibility}: Adding terms with zero frequency does not change the entropy, allowing for consistent comparison of portfolios with different numbers of distinct terms \citep{rao2004quadratic}. This property maintains the integrity of the diversification measure even if some assets have no exposure to specific types of information.
\item \textbf{Scale invariance}: Scaling the weights by a constant does not impact the LR metric value. This is particularly useful since, in portfolio optimization, we often require the weights of the assets to sum to 1. A fraction of a text is meaningless. However, by scaling the weights by a large number, we will have the LR metric unchanged, and we can examine the combination of texts (the proof of the scale invariance property is provided in Appendix~\ref{app:proof}).
\end{enumerate}
These properties make the lexical ratio a robust and theoretically sound measure of portfolio diversification based on the lexical content of asset-related documents. It captures the notion of diversification in a way that aligns with intuitive understanding while providing a mathematically rigorous framework for portfolio optimization.

\subsection{Examples of LR calculation}

To illustrate the application of the LR measure, we present two examples to provide an intuitive understanding of how portfolios constructed using LR are well-diversified. In these examples, we have a total word collection of $m$ = $\{$T, H, O, E$\}$ for normalization, where T is a word relating to the technology sector, H is a word relating to the healthcare sector, E is a word relating to the energy sector, and O is a word relating to oil shipment news.\footnote{The weights in the examples are scaled by 10 for the combined text for illustration purposes. Given that LR is scale-invariant, this scaling does not affect the numeric values, and this number is arbitrary.}

\subsubsection{Balancing technology and healthcare assets}

In this first example, we seek to illustrate how LR can be used to measure the number of different sectors in a portfolio. Consider a portfolio with two assets representing different combinations of technology (T) and healthcare (H) assets:
\begin{itemize}
\setlength\itemsep{0em}
\item Asset 1: T (i.e., a technology-related asset).
\item Asset 2: H (i.e., a health-related asset).
\end{itemize}
Let us assume the following weights associated with Assets 1 and 2: $w_1 = 0.5$ and $w_2 = 0.5$. Let us further assume the combined string representation
\begin{equation}
    \text{``TTTTTHHHHH''} . \notag
\end{equation}
Then, LR is computed as
$$
\text{LR} = -\frac{1}{\log(4)} \left(\frac{5}{10} \log\left(\frac{5}{10}\right) + \frac{5}{10} \log\left(\frac{5}{10}\right)\right) = \frac{1}{\log(4)}. 
$$
because we have $m = 4$ words. This indicates a balanced portfolio containing multiple assets with balanced weights.

Alternatively, let us assume the weights $w_1 = 0$ and $w_2 = 1$, and the combined string representation
\begin{equation}
    \text{``HHHHHHHHHH''} . \notag
\end{equation}
Then, LR is computed as
$$
\text{LR} = -\frac{1}{\log(4)} \left(\frac{10}{10} \log\left(\frac{10}{10}\right) + \frac{10}{10} \log\left(\frac{10}{10}\right)\right) = 0,
$$
which indicates a less diversified portfolio of assets from only one sector.

\subsubsection{Balanced distribution based on news effects}

In this second example, we seek to illustrate how LR can be used to measure how balanced a portfolio is about news. For example, we will be considering assets affected to varying degrees by news regarding oil shipments in the Persian Gulf. This news can highly affect certain assets while having little impact on others. For example, energy sector assets will likely be more affected than real estate sector assets. An asset with ties to the Middle East will likely suffer more than one without such ties. It is important to have a balanced and diverse portfolio that would protect our investments from losses resulting from the negative impacts of such news while also allowing us to leverage the advantages that assets could have. A portfolio of assets highly impacted by oil shipment news is vulnerable to negative news. In contrast, a portfolio of many assets not significantly impacted by such news could prevent an investor from taking advantage of relevant opportunities such as oil price increases. Consider a portfolio with four assets representing different combinations of words:
\begin{itemize}
\setlength\itemsep{0em}
\item Asset 1: T (i.e., technology-related asset).
\item Asset 2: EN (i.e., energy-related asset that is slightly impacted by oil shipping news).
\item Asset 3: ENN (i.e., energy-related asset that is moderately impacted by oil shipping news).
\item Asset 4: ENNN (i.e., energy-related asset that is highly impacted by oil shipping news).
\end{itemize}
Alternatively, let us assume the weights $w_1 = 0.4$, $w_2 = 0.3$, $w_3 = 0.2$, and $w_4 = 0.1$, and the combined string representation
\begin{equation}
    \text{``TTTTENENENENNENNENNN''} . \notag
\end{equation}
Then, LR is computed as
$$
\text{LR} = - \frac{13}{\log(4)} \left(\frac{10}{20} \log\left(\frac{10}{20}\right) + \frac{4}{20} \log\left(\frac{4}{20}\right)  + \frac{6}{20} \log\left(\frac{6}{20}\right)\right) = 0.743.
$$
This portfolio has a balanced distribution of assets that are less affected by oil shipment news and those less affected by it. In cases where such news results in higher returns, we ensure that we can use this to our advantage while protecting our portfolio against cases where such news might be negative through a balanced construction of assets.

Alternatively, let us assume the weights  $w_1 = 0$, $w_2 = 0$, $w_3 = 0$, and $w_4 = 1$, and the combined string representation
\begin{equation}
    \text{``ENNNENNNENNNENNNENNNENNNENNNENNNENNNENNN''} . \notag
\end{equation}
Then, LR is computed as
$$
\text{LR} = - \frac{13}{\log(4)}\left(\frac{10}{40} \log\left(\frac{10}{40}\right) + \frac{30}{40} \log\left(\frac{30}{40}\right)\right) = 0.405.
$$
It can be seen that the portfolio that is more affected by oil shipping news is less diverse than the one that is affected in a balanced manner based on the lower LR metric.

\section{Comparison and analyses\label{sec:comparison}}

\citet{lin2022diversification} introduced an axiomatic framework for constructing portfolio diversification metrics. While our measure adheres to several of their proposed properties, such as scale invariance, non-negativity, and normalization, we have opted for an experimental approach in this study. Although axiomatic methods provide a strong theoretical foundation, they can sometimes be overly rigid and may not fully capture real-world dynamics. For instance, constraints like convexity or specific definitions of rationality can impose unnecessary limitations. Furthermore, the reliance on risk measures produces metrics that focus on extreme data, which might weaken their performance. Since our primary focus is achieving robust performance in practical scenarios, we believe an experimental methodology better suits our objectives. To this end, we have designed several real-world experiments.

Since LR uses lexical information and is an unconventional metric, we first examine its relationship to conventional return-based metrics. This is done through a correlation and regression analysis of various portfolios and conditional non-parametric methods. Furthermore, we examine the robustness of these metrics through a random-weight analysis. Finally, we test out-of-sample and optimize real-world portfolios using these metrics. We are using Markowitz's volatility (i.e., the standard deviation of the asset of the portfolio),  a diversification ratio (DR) based on the standard deviation, and a DR based on value-at-risk (VaR) in all experiments to examine how LR relates to conventional and widely accepted metrics and to show the drawbacks of relying on numeric asset return data alone.

The diversification ratio based on the standard deviation, introduced by \citet{choueifaty2008toward}, is a widely recognized measure of portfolio diversification. The measure DR$_{\text{SD}}$ is defined as the ratio of the weighted average volatility of the portfolio's assets to the overall portfolio volatility:
\begin{equation}
\text{DR}_{\text{SD}}(\pmb{w}) = \frac{\pmb{w}^{\top} \text{diag} \left( \mathrm{Var}[\pmb{X}]\right)}{\sqrt{\pmb{w}^{\top} \mathrm{Var}[\pmb{X}] \pmb{w}}} ,
\end{equation}
where $\pmb{w} = [ \,\, w_1 \quad w_2 \quad ... \quad w_n \,\,]^{\top}$ is the vector of portfolio weights, $\pmb{X} = [ \,\, X_1 \quad X_2 \quad ... \quad X_n \,\,]$ is the vector of asset returns, and $\mathrm{Var}[\pmb{X}]$ is the covariance matrix of asset returns.

Other formulations of DR using risk measures have been explored in studies by \cite{degen2010risk} and \cite{embrechts2009multivariate}. We also use the diversification ratio (DR) based on the VaR, which quantifies the risk reduction achieved through diversification to examine the uses of risk-based diversification measures. The formulation is given by:
\begin{equation}
\text{DR}_{\text{VaR}}(\pmb{w}) = \frac{\text{VaR}_{\alpha}\left(\pmb{w}^{\top} \pmb{X} \right)}{\sum_{i=1}^{n} w_i \text{VaR}_{\alpha}(X_i)} ,
\end{equation}
where $\text{VaR}_{\alpha}(X_i)$ represents the VaR at confidence level $\alpha$ for asset return $X_i$. A higher DR indicates less diversification, implying that the assets are highly correlated and the portfolio does not benefit significantly from diversification. Conversely, a lower DR suggests greater diversification, as the portfolio VaR is significantly less than the sum of individual VaRs, indicating effective risk reduction through diversification \cite{Jorion2007, Litterman2003}. We use a level of $\alpha=0.05$ in all cases.

We rely on 19 real-world portfolios for all of our experiments. We use sector-specific, random mixed-sector portfolios and portfolios with varying levels of volatility. The details are outlined below.

The S\&P 500 stocks are divided into 11 sectors according to the Global Industry Classification Standard (GICS). We selected the top 10 companies for each sector based on market capitalization on May 5, 2023. This leads to 11 portfolios---one for each sector: information technology, health care, financials, consumer discretionary, communication services, industrials, consumer staples, energy, utilities, materials, and real estate.\footnote{The sectors' respective top 10 companies, given by their tickers, are as follows: AAPL, MSFT, NVDA, AVGO, ADBE, CRM, CSCO, ORCL, INTC, QCOM for information technology, UNH, JNJ, LLY, PFE, MRK, ABBV, TMO, ABT, DHR, BMY for health care, BRK.B, JPM, BAC, WFC, MS, GS, C, BLK, AXP, SCHW for financials, AMZN, TSLA, HD, MCD, NKE, LOW, SBUX, BKNG, TGT, TJX for consumer discretionary, GOOGL, GOOG, META, CMCSA, VZ, DIS, T, NFLX, TMUS, CHTR for communication services,  RTX, UNP, HON, LMT, GE, BA, CAT, MMM, NOC, DE for industrials, PG, PEP, KO, WMT, COST, PM, MDLZ, CL, KMB, MO for consumer staples, XOM, CVX, COP, SLB, EOG, MPC, PSX, OXY, VLO, PXD for energy,  EE, DUK, SO, D, AEP, EXC, SRE, ED, XEL, PEG for utilities, LIN, APD, SHW, DD, NEM, FCX, ECL, PPG, IP, VMC for materials, and AMT, PLD, CCI, EQIX, PSA, DLR, SPG, O, AVB, WELL for real estate.}

Five random mixed portfolios are also constructed. We build two of those by selecting two companies from each of the 11 sectors, ensuring the portfolio has stocks from each sector to be used as an example of a diverse portfolio. The other three mixed portfolios are built by taking 20 stocks randomly from all of the S\&P 500 stocks without any constraints, including examples of what a portfolio manager might want to analyze for a client.\footnote{The mixed portfolios are given as follows: AAPL, MSFT, UNH, JNJ, BRK.B, JPM, AMZN, TSLA, GOOGL, META, RTX, UNP, PG, PEP, XOM, CVX, NEE, DUK, LIN, APD, AMT, PLD for the random mixed portfolio 1, NVDA, ADBE, PFE, MRK, BAC, WFC, HD, MCD, CMCSA, NFLX, HON, LMT, WMT, KO, COP, SLB, SO, EXC, DD, FCX, CCI, EQIX for the random mixed portfolio 2, TFC, APD, EXPD, EMN, CHD, ED, AMGN, MSCI, FTNT, VST, TGT, MKC, PNW, AXON, NDAQ, NCLH, SMCI, FITB, DVA, TYL for the random mixed portfolio 3, AJG, MSCI, COF, AMCR, XOM, SRE, NSC, FCX, ZTS, MKTX, COO, HON, CB, PH, NWSA, IR, WM, GEV, SHW, ORCL for the random mixed portfolio 4, and  ETN, FIS, SPGI, AIZ, PNC, ADI, PODD, TMO, UNH, DOV, FITB, SNPS, SLB, ETR, HRL, FTNT, IT, NRG, IQV, BWA for the random mixed portfolio 5.}

Three portfolios were constructed based on stock volatility to examine the performances of diversification metrics in various scenarios. High-volatility stocks are selected from sectors known for significant price fluctuations, while low-volatility stocks were chosen from more stable sectors. The high-volatility stocks are identified based on their historical price fluctuations and beta values, sourced from financial analysis platforms such as \citet{investopedia1, investopedia2}, \citet{fool}, and \citet{nasdaq}. Low-volatility stocks are chosen based on their stability and lower beta values, with data sourced from \citet{spglobal} and \citet{rbc}. We also consider a portfolio that includes both low- and high-volatility stocks called the mixed volatility portfolio.\footnote{The portfolios with varying levels of volatility are defined as follows: TSLA, NVDA, AMZN, META, NFLX, AMD, MU, LRCX for the high-volatility portfolio, JNJ, PG, KO, PEP, NEE, DUK, SO, EXC for the low-volatility portfolio, and TSLA, NVDA, AMZN, META, NFLX, JNJ, PG, KO, PEP, NEE for the mixed volatility portfolio.}

\subsection{Relationship analysis}

This section explores the relationship between LR and traditional portfolio diversification metrics, specifically Markowitz's volatility, DR$_{\text{SD}}$, and DR$_{\text{VaR}}$. The analysis is divided into two parts: (i) a nonparametric conditional dependence analysis to capture the evolving relationship between LR and traditional metrics over time for a fixed equal weights portfolio, and (ii) a regression and correlation analysis that examines how varying portfolio weights impact the relationship between LR and traditional metrics. Together, these analyses provide insights into how LR interacts with traditional diversification measures across both fixed and flexible portfolio configurations. While LR shows significant relationships with established metrics, it provides unique insights and does not entirely overlap.

\subsubsection{Nonparametric conditional dependence analysis}

The primary objective of this analysis is to capture how the relationship between LR and traditional diversification metrics evolves for a fixed portfolio conditioned on the VIX---a proxy for market volatility derived from S\&P 500 index options. To measure this relationship, we employ the conditional dependence coefficient \(T_n\) as proposed by \cite{azadkia}; it is defined as
\[
T_n = \frac{\sum_{i=1}^{n} \left(\min\{R_i, R_{M(i)}\} - \min\{R_i, R_{N(i)}\}\right)}{\sum_{i=1}^{n} \left(R_i - \min\{R_i, R_{N(i)}\}\right)},
\]
where $n$ is the number of time-frame windows analyzed. \(R_i\) represents the rank of the $i^{\text{th}}$ observation of LR or the traditional metrics conditioned on the VIX, \(M(i)\) is the nearest neighbour of the $i^{\text{th}}$ observation based on LR or the traditional metrics, and \(N(i)\) is the nearest neighbour based only on LR.

The coefficient \(T_n\) ranges from 0 (indicating independence) to 1 (indicating strong dependence). Higher values of \(T_n\) suggest a stronger relationship between LR and traditional metrics. In this context, the coefficient captures how similar the rankings of LR and traditional metrics are when conditioned on the VIX. If the rankings are closely aligned, \(T_n\) is high, indicating a strong relationship. Conversely, if the rankings diverge, \(T_n\) is lower, indicating weaker dependence.

The analysis uses a rolling window technique, where each window covers 180 days of data (six months) and shifts forward by 90 days (three months). This approach allows us to track how the dependence between LR and traditional metrics changes over time, while conditioning on market volatility. 

\begin{table}[ht!]
{\fontsize{9.5}{10}\selectfont
\centering
\topcaption{\textbf{{Azadkia--Chatterjee conditional dependence measures between LR and traditional diversification measures}\label{conditional}}}
\begin{tabularx}{\linewidth}{X*{3}{d{8.9}}} 
\toprule
\textbf{Portfolio} &  \multicolumn{1}{c}{\textbf{Volatility}} & \multicolumn{1}{c}{$\text{\textbf{DR}}_{\text{\textbf{SD}}}$} & \multicolumn{1}{c}{$\text{\textbf{DR}}_{\text{\textbf{VaR}}}$} \\
\cmidrule(lr){1-1} \cmidrule(lr){2-2} \cmidrule(lr){3-3} \cmidrule(lr){4-4} 
Information technology              & 0.41 & 0.51 & 0.00 \\
Health care             & 0.19 & 0.16 & 0.28 \\
Financials              & 0.22 & 0.49 & 0.41 \\
Consumer discretionary  & 0.30 & 0.43 & 0.14 \\
Communication services  & 0.30 & 0.43 & 0.30 \\
Industrials             & 0.00 & 0.00 & 0.00 \\
Consumer staples        & 0.11 & 0.69 & 0.51 \\
Energy                  & 0.35 & 0.03 & 0.13 \\
Utilities               & 0.00 & 0.07 & 0.21 \\
Materials               & 0.00 & 0.20 & 0.07 \\
Real estate             & 0.31 & 0.17 & 0.10 \\[1ex]
Random mixed portfolio 1       & 0.06 & 0.28 & 0.00 \\
Random mixed portfolio 2       & 0.30 & 0.32 & 0.27 \\
Random mixed portfolio 3       & 0.00 & 0.03 & 0.00 \\
Random mixed portfolio 4       & 0.00 & 0.08 & 0.22 \\
Random mixed portfolio 5       & 0.00 & 0.00 & 0.00 \\[1ex]
High volatility                & 0.47 & 0.47 & 0.47 \\
Low volatility                 & 0.00 & 0.00 & 0.14 \\
Mixed volatility               & 0.36 & 0.42 & 0.44 \\
\bottomrule
\end{tabularx}
}
\end{table}
\normalsize

Table \ref{conditional} reveals varying dependence between LR and traditional metrics. For example, the consumer staples portfolio demonstrates a strong relationship between LR and DR$_{\text{SD}}$ (i.e., \(T_n = 0.69\)), while the industrials portfolio shows no dependence across all metrics. These results highlight that LR and traditional metrics have meaningful but diverse relationships when examined for an equal-weight portfolio over time. This means that they share information while differing meaningfully in their overall assessment.

\subsubsection{Regression and correlation analysis}

In addition to the nonparametric analysis, we perform regressions and correlation analyses to assess how varying portfolio weights impact the relationship between LR and traditional metrics. Random-weight portfolios were generated using a Dirichlet distribution, ensuring that all portfolios are non-negative and sum to one. The Dirichlet distribution is defined as
\[
f(w_1, w_2, ..., w_K; \alpha_1, \alpha_2, \ldots, \alpha_K) = \frac{1}{B(\pmb{\alpha})} \prod_{i=1}^{K} w_i^{\alpha_i - 1},
\]
where \(w_i \geq 0\), \(\sum_{i=1}^{K} w_i = 1\), and \(B(\pmb{\alpha})\) is the multivariate Beta function. This distribution is well-suited for generating realistic portfolio weights. We generated 1000 random weights for each portfolio in our experiments.

For each of the three traditional metrics, we conduct regression tests and Pearson correlation calculations for all 19 portfolios over a period from January 1, 2018, to June 30, 2024, using a 90-day rolling window described above, and each analysis is performed on data spanning 180 days (six months). This yields a total of 969 regression tests for 19 portfolios based on the following regression model:
\[
\text{Metric}_t = \beta_0 + \beta_1 \, \text{LR}_t + \epsilon_t,
\]
where \(\text{Metric}_t\) represents either Markowitz's volatility, DR$_{\text{SD}}$, or DR$_{\text{VaR}}$, and $\beta_0$ and \(\beta_1\) are the regression coefficients capturing the intercept and the relationship between LR and the traditional metric, respectively.

Statistical significance is evaluated at the 5\% level. Approximately 98\% of the regressions are significant, indicating robust relationships between LR and traditional metrics---a very high figure. 

\begin{figure}[h]
    \centering
    \includegraphics[width=0.8\textwidth]{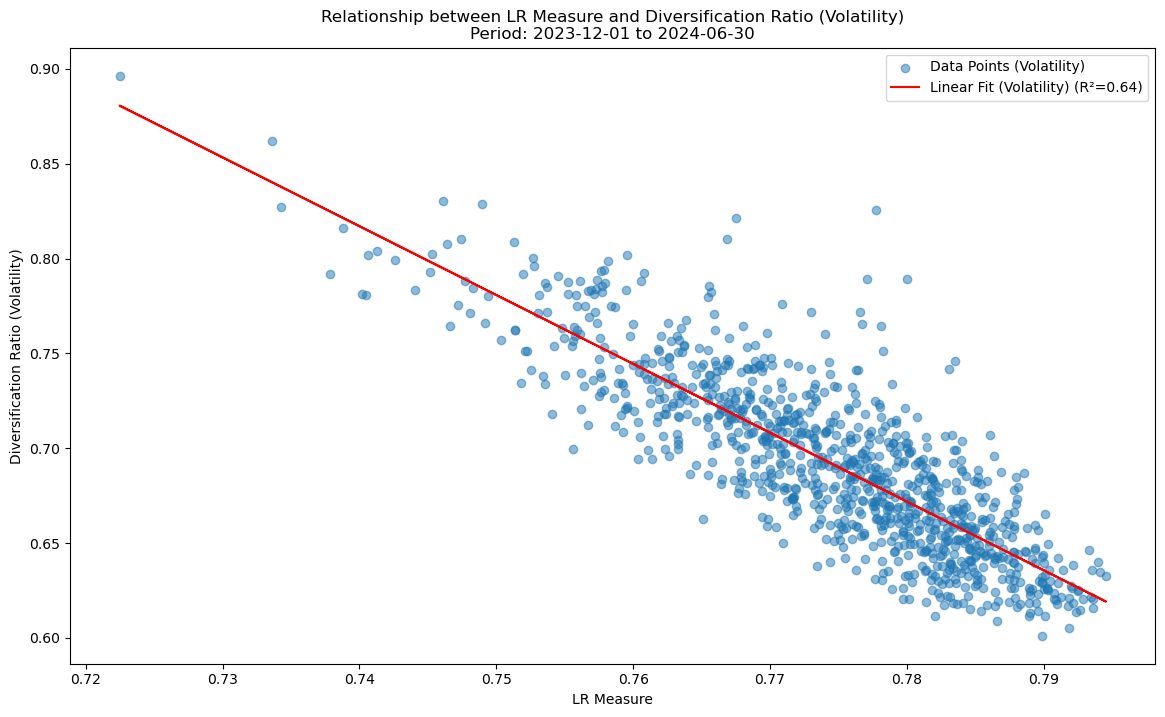}
    \caption{\textbf{Relationship between LR and the diversification ratio based on the standard deviation for Random mixed portfolio 3}}
    \label{fig:corr}
\end{figure}

Pearson correlation coefficients are also computed for each of the 969 analyses to measure the degree of correlation between LR and traditional metrics. These results are summarized in Table~\ref{meancorrelation}. The correlations were generally negative, suggesting that traditional metrics tend to decrease as LR increases (indicating higher diversification). This negative relationship aligns with expectations, as higher diversification typically results in lower values for traditional metrics such as volatility. An example of the relationships found is illustrated in Figure~\ref{fig:corr}.

\begin{table}[ht!]
{\fontsize{9.5}{10}\selectfont
\centering
\topcaption{\textbf{{Average correlation between LR and traditional measures}\label{meancorrelation}}}
\begin{tabularx}{\linewidth}{X*{3}{d{8.9}}} 
\toprule
\textbf{Portfolio} & \multicolumn{1}{c}{\textbf{Volatility}} & \multicolumn{1}{c}{$\text{\textbf{DR}}_{\text{\textbf{SD}}}$} & \multicolumn{1}{c}{$\text{\textbf{DR}}_{\text{\textbf{VaR}}}$} \\
\cmidrule(lr){1-1} \cmidrule(lr){2-2} \cmidrule(lr){3-3} \cmidrule(lr){4-4} 
Information technology        & -0.5141 & -0.4950 & -0.2861 \\ 
Health care      & -0.2864 & -0.4143 & -0.1673 \\ 
Financials        & -0.2593 & -0.2344 & -0.1285 \\ 
Consumer discretionary & -0.1586 & -0.3975 & -0.1219 \\ 
Communication services & -0.3016 & -0.3123 & -0.2385 \\
Industrials & -0.0391 & -0.3198 & -0.1585 \\
Consumer staples  & -0.2271 & -0.2219 & -0.1553 \\ 
Energy            & -0.3732 & -0.2323 & -0.0211 \\ 
Utilities         & -0.4612 & -0.1032 & -0.1052 \\ 
Materials         & -0.3092 & -0.5342 & -0.3925 \\ 
Real estate      &  0.1037 & -0.3619 & -0.1896 \\[1ex]
Random mixed portfolio 1 & -0.2894 & -0.2859 & -0.0765 \\ 
Random mixed portfolio 2 & -0.1509 & -0.2857 & -0.2388 \\ 
Random mixed portfolio 3 & -0.3237 & -0.3202 & -0.0785 \\ 
Random mixed portfolio 4 & -0.1481 & -0.2605 & -0.1216 \\ 
Random mixed portfolio 5 & -0.2420 & -0.2306 & -0.0536 \\[1ex]
High volatility         & -0.4275 & -0.5270 & -0.1708 \\ 
Low volatility          & -0.4872 & -0.3488 & -0.2540 \\ 
Mixed volatility        & -0.2153 & -0.2988 & -0.0942 \\ 
\bottomrule
\end{tabularx}
}
\end{table}
\normalsize

\subsection{Robustness testing of portfolio diversification metrics}

This section evaluates the robustness of portfolio metrics using the coefficient of variation (CV), highlighting that LR is a reliable diversification measure. The CV is used to assess the relative variability of metrics such as LR, Markowitz's volatility, DR$_{\text{SD}}$, and DR$_{\text{VaR}}$ across time. This analysis provides insights into how consistently these metrics perform in different market conditions. CV is defined as
\[
CV = \frac{\sqrt{\mathrm{Var}[\text{Metric}_t]}}{\mathrm{E}[\text{Metric}_t]}
\]
where $\mathrm{Var}[\text{Metric}_t]$ represents the variance of the metric and $\mathrm{E}[\text{Metric}_t]$ is its mean. The CV allows for a normalized comparison of the relative stability of different metrics. Lower CV values indicate greater stability and consistency over time, while higher CV values suggest increased sensitivity to fluctuations in market conditions.

For each portfolio, we generate 1000 random weight combinations using the Dirichlet distribution to account for various weight configurations. To conduct this analysis, we employ a rolling window approach with a window size of 180 days and a step size of 90 days; the analysis covers the period from January 1, 2018, to June 30, 2024. For each window, the diversification metrics are calculated across various portfolios. After computing these metrics, each CV is determined to assess their stability across time.

Table~\ref{tab:cv_results} presents the mean results, where LR consistently exhibits lower CV values across all portfolios. This suggests that LR is a more stable and reliable diversification measure compared to traditional metrics like volatility and diversification ratios. The higher CV values for other metrics indicate greater variation and less reliability.

\begin{table}[ht!]
{\fontsize{9.5}{10}\selectfont
\centering
\topcaption{\textbf{{Coefficient of variation for LR and traditional measures}\label{tab:cv_results}}}
\begin{tabularx}{\linewidth}{X*{4}{d{6.9}}} 
\toprule
\textbf{Portfolio} & \multicolumn{1}{c}{\textbf{LR}} & \multicolumn{1}{c}{\textbf{Volatility}} & \multicolumn{1}{c}{$\text{\textbf{DR}}_{\text{\textbf{SD}}}$} & \multicolumn{1}{c}{$\text{\textbf{DR}}_{\text{\textbf{VaR}}}$} \\
\cmidrule(lr){1-1} \cmidrule(lr){2-2} \cmidrule(lr){3-3} \cmidrule(lr){4-4} \cmidrule(lr){5-5} 
Information technology                     & 0.0594            & 0.3125                           & 0.0854         & 0.3134          \\ 
Health care                    & 0.0558            & 0.3840                           & 0.1349         & 0.3853          \\ 
Financials                     & 0.0516            & 0.4550                           & 0.0646         & 0.4469          \\ 
Consumer discretionary         & 0.0628            & 0.4007                           & 0.0935         & 0.4264          \\ 
Communication services         & 0.0646            & 0.3574                           & 0.1222         & 0.3929          \\ 
Industrials                    & 0.0505            & 0.4210                           & 0.1102         & 0.3944          \\ 
Consumer staples               & 0.0528            & 0.4209                           & 0.0941         & 0.3732          \\ 
Energy                         & 0.0519            & 0.4741                           & 0.0528         & 0.4633          \\ 
Utilities                      & 0.0587            & 0.4647                           & 0.0587         & 0.4723          \\ 
Materials                      & 0.0600            & 0.3811                           & 0.0976         & 0.3514          \\ 
Real estate                    & 0.0554            & 0.4816                           & 0.0828         & 0.4730          \\[1ex]
Random mixed portfolio 1             & 0.0463            & 0.4864                           & 0.1624         & 0.4870          \\ 
Random mixed portfolio 2             & 0.0480            & 0.5225                           & 0.1697         & 0.5149          \\ 
Random mixed portfolio 3             & 0.0327            & 0.4906                           & 0.1431         & 0.4650          \\ 
Random mixed portfolio 4             & 0.0423            & 0.5311                           & 0.1451         & 0.5132          \\ 
Random mixed portfolio 5             & 0.0499            & 0.4200                           & 0.1482         & 0.3952          \\[1ex]
High volatility                & 0.0588            & 0.2610                           & 0.0728         & 0.3083          \\
Low volatility                 & 0.0550            & 0.5508                           & 0.0761         & 0.5674          \\ 
Mixed volatility               & 0.0683            & 0.3630                           & 0.1123         & 0.3817          \\ 
\bottomrule
\end{tabularx}
}
\end{table}
\normalsize

\subsection{Real-world optimization testing}

We use news headlines from the Unicorn Data Services Historic Data (EODHD) for each ticker to calculate LR. The data are from January 1, 2018, to June 30, 2024. This time frame is chosen based on the availability of organized textual data in various packages and datasets. The portfolios are optimized on windows of 24 months (720 days) and tested on the consecutive six months (180 days), creating an 80-20 optimization-testing split for out-of-sample testing. The window is rolled every three months (90 days). This makes 17 total experiments for each portfolio. The word collection used for LR calculations is all words available in the headlines for the selected time frame. The mean results of the 17 windows for each portfolio are presented in Table~\ref{tab:results}.

\begin{figure}[ht!]
    \centering
    \includegraphics[width=0.63\textwidth]{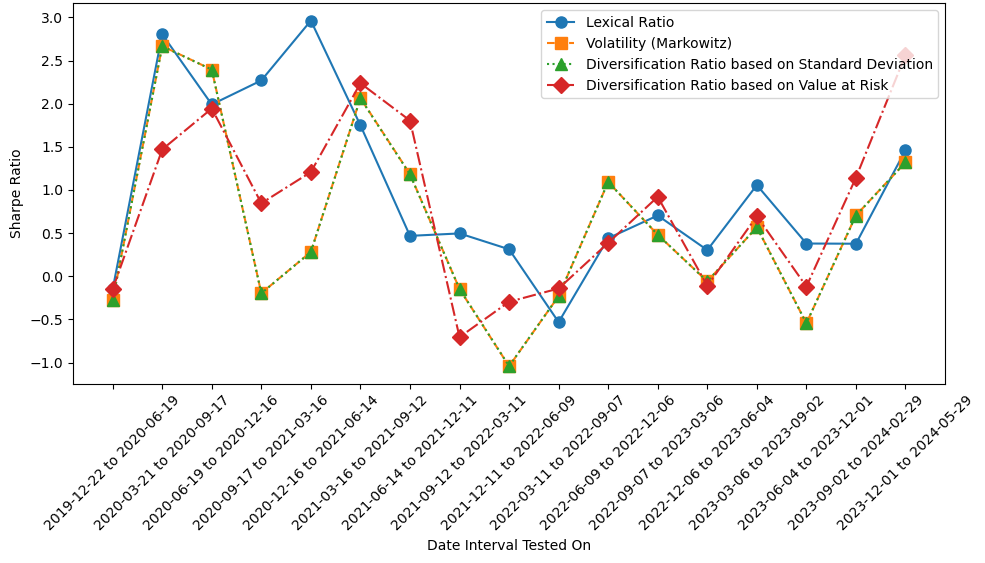}
    \caption{\textbf{The time series of the Sharpe ratio for Random mixed portfolio 1 over the testing window}}
    \label{fig:sharpe}
\end{figure}

\begin{figure}[ht!]
    \centering
    \includegraphics[width=0.63\textwidth]{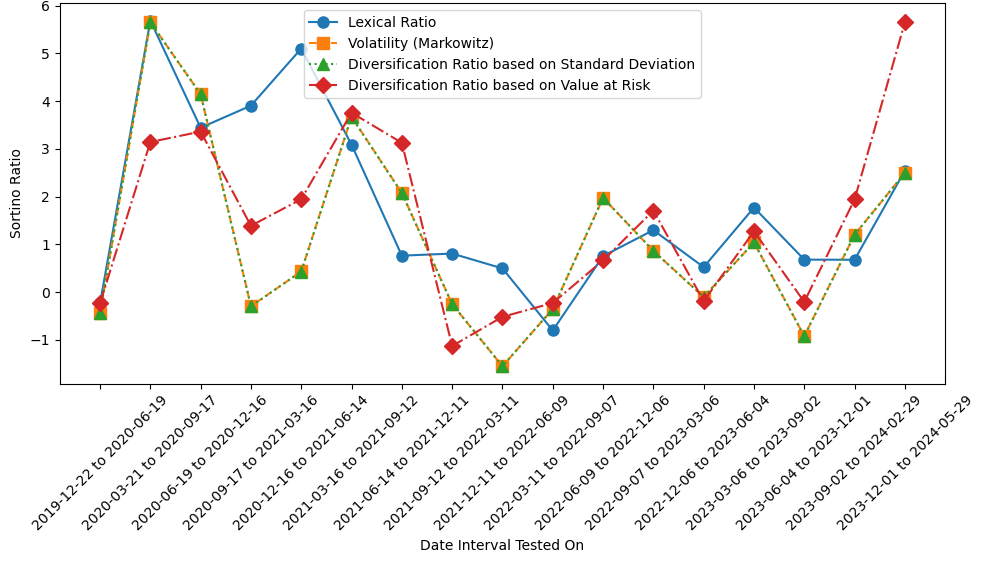}
    \caption{\textbf{The time series of the Sortino ratio for Random mixed portfolio 1 over the testing window}}
    \label{fig:sortino}
\end{figure}

We use the sequential least-squares quadratic programming (SLSQP) method, as discussed by \cite{boggs2000sequential}, for all of our optimization, which is particularly effective for this task due to its ability to handle complex constraints efficiently. SLSQP is a gradient-based optimization technique that excels in problems where the objective function is smooth and differentiable, making it suitable for portfolio optimization scenarios that require both equality and inequality constraints to be satisfied. The return constraints for the optimization process are set to be greater than or equal to \{0.07, 0.1, 0.13, 0.16\} of annual return to account for a diverse set of return constraints. We optimize for each constraint and take the mean of the optimization of these constraints. Our objective function includes (the negative of) LR, Markowitz's volatility, the diversification ratio based on the standard deviation (DR$_{\text{SD}}$), the diversification ratio based on VaR (DR$_{\text{VaR}}$).\footnote{We use the negative of LR because as low LRs indicate less diversification.} The initial weights used in the optimization are drawn randomly from a Dirichlet distribution. The risk-free interest rate is set to 0.024 based on the average 10-year US bond yield over the period.

\begin{figure}[ht!]
    \centering
    \includegraphics[width=0.63\textwidth]{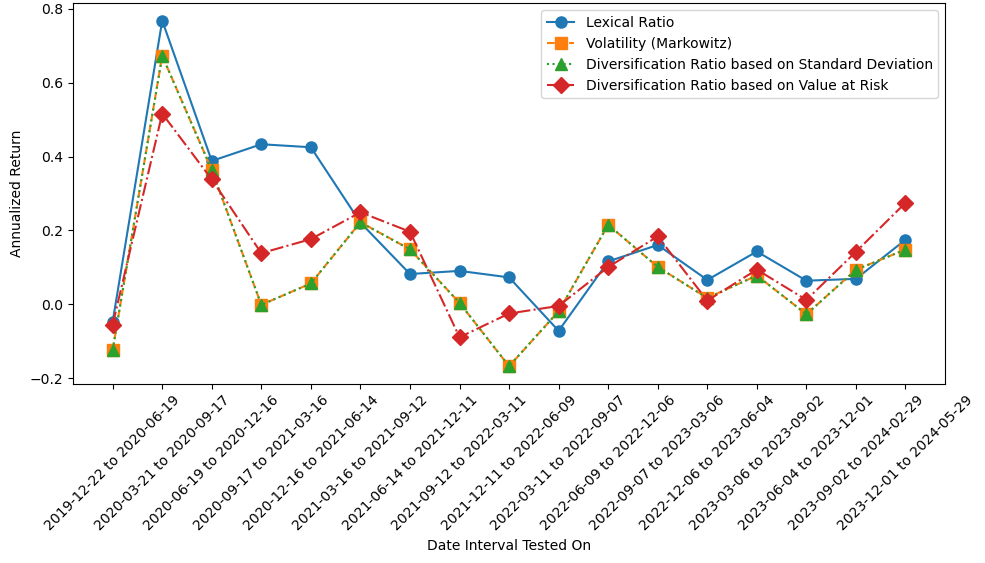}
    \caption{\textbf{The time series of the annualized returns for Random mixed portfolio 1 over the testing window}}
    \label{fig:returns}
\end{figure}

\begin{figure}[ht!]
    \centering
    \includegraphics[width=0.65\textwidth]{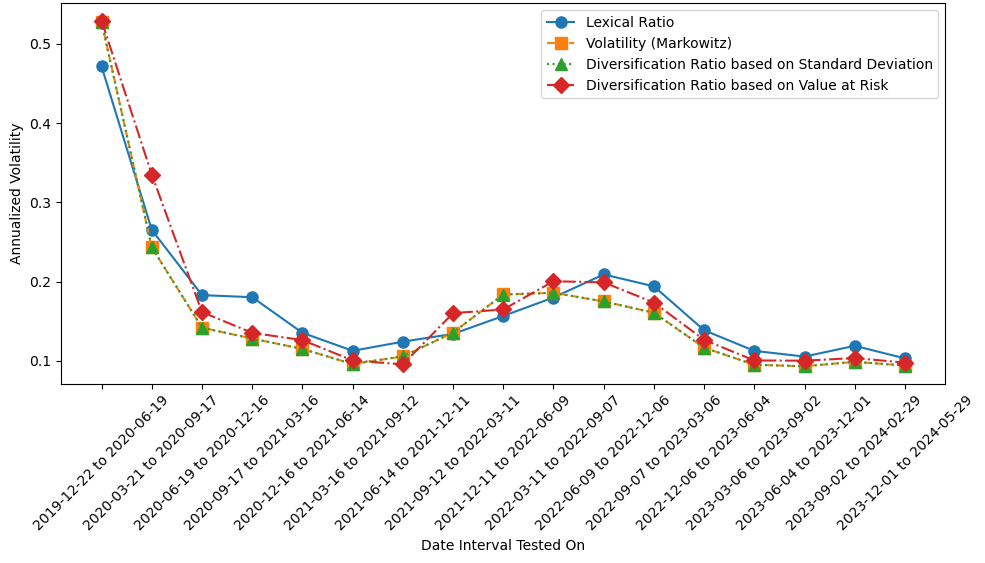}
    \caption{\textbf{The time series of the annualized volatility for Random mixed portfolio 1 over the testing window}}
    \label{fig:vol}
\end{figure}

LR significantly outperforms existing measures in terms of its Sharpe ratio and Sortino ratio, meaning that it offers the best return to risk among these methods. This highlights the robustness of LR and its suitability for use in real-world applications. This can be attributed to the robustness of concentrating large volumes of information in a single metric instead of relying only on historical return data.

\begin{figure}[ht!]
    \centering
    \includegraphics[width=0.63\textwidth]{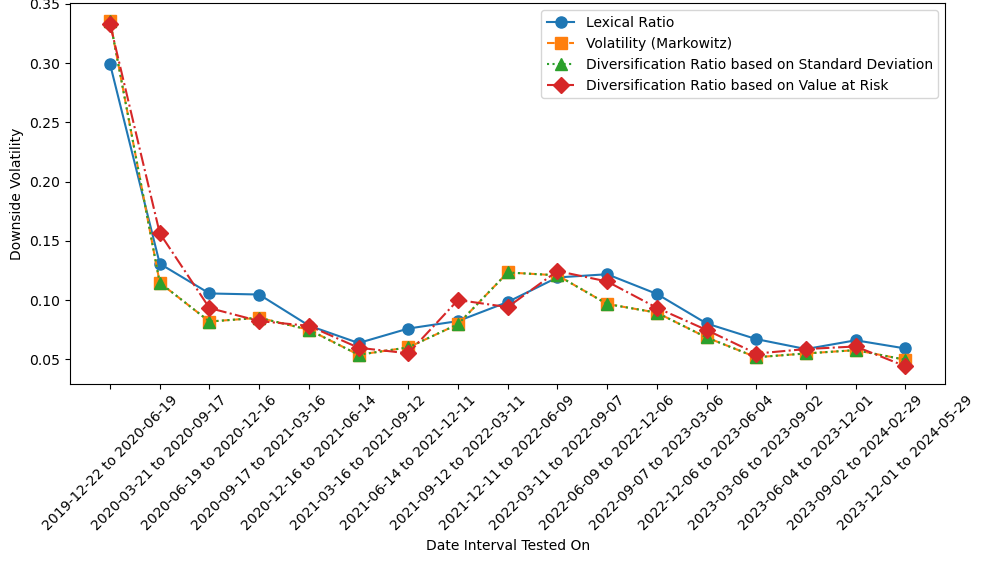}
    \caption{\textbf{The time series of the downside volatility for Random mixed portfolio 1 over the testing window}}
    \label{fig:downside}
\end{figure}

Figures~\ref{fig:sharpe} to \ref{fig:downside} display how the Sharpe ratio, the Sortino ratio, the annualized returns, the annualized volatility, and the downside volatility change over the testing window for Random mixed portfolio 1, respectively. The other portfolios displayed similar trends, all being characterized by substantial variability and fluctuations in the results of traditional diversification metrics. This behavior contrasts with the more stable patterns observed in the Lexical Ratio (LR) graphs, which consistently achieve higher returns and exhibit lower volatility compared to traditional metrics. Notably, the LR shows a more linear progression in both the Sharpe and Sortino ratio graphs, suggesting a more robust performance. All metrics experienced a downturn during the COVID-19 period, indicating the vulnerability to systematic market shocks. However, following this period, each method demonstrated a significant recovery in performance.

\begin{table}[ht!]
{\fontsize{9.5}{10}\selectfont
\centering
\topcaption{\textbf{{Average metric across different portfolios constructed using LR, volatility, diversification ratio based on SD, and diversification based on VaR}\label{tab:results}}}
\begin{tabularx}{\linewidth}{XX*{4}{d{3.6}}} 
\toprule
\textbf{Portfolio} & \textbf{Metric} & \multicolumn{1}{c}{\textbf{LR}} & \multicolumn{1}{c}{\textbf{Volatility}} & \multicolumn{1}{c}{$\text{\textbf{DR}}_{\text{\textbf{SD}}}$} & \multicolumn{1}{c}{$\text{\textbf{DR}}_{\text{\textbf{VaR}}}$} \\
\cmidrule(lr){1-1} \cmidrule(lr){2-2} \cmidrule(lr){3-3} \cmidrule(lr){4-4} \cmidrule(lr){5-5} \cmidrule(lr){6-6} 
\multirow{5}{*}{\textbf{Information technology}} & Sharpe ratio       & 1.5041 & 1.1116 & 1.1117 & 1.1013 \\
& Sortino ratio      & 2.9977 & 2.2570 & 2.2572 & 2.1737 \\
& Annualized returns  & 0.4139 & 0.2376 & 0.2376 & 0.2513 \\
& Annualized volatility & 0.3024 & 0.2316 & 0.2316 & 0.2419 \\
& Downside volatility  & 0.1669 & 0.1306 & 0.1306 & 0.1370 \\[1ex]
\multirow{5}{*}{\textbf{Health care}} & Sharpe ratio       & 0.9823 & 0.6283 & 0.6287 & 0.7736 \\
& Sortino ratio      & 1.8833 & 1.1939 & 1.1944 & 1.4689 \\
& Annualized returns  & 0.1918 & 0.1310 & 0.1311 & 0.1557 \\
& Annualized volatility & 0.1776 & 0.1620 & 0.1620 & 0.1725 \\
& Downside volatility  & 0.0986 & 0.0924 & 0.0924 & 0.0968 \\[1ex]
\multirow{5}{*}{\textbf{Financials}} & Sharpe ratio       & 0.9614 & 0.6903 & 0.6903 & 0.8336 \\
& Sortino ratio      & 1.9404 & 1.2663 & 1.2664 & 1.6158 \\
& Annualized returns  & 0.2510 & 0.1603 & 0.1604 & 0.1966 \\
& Annualized volatility & 0.2582 & 0.2416 & 0.2416 & 0.2410 \\
& Downside volatility  & 0.1429 & 0.1377 & 0.1377 & 0.1368 \\[1ex]
\multirow{5}{*}{\textbf{Consumer discretionary}} & Sharpe ratio       & 0.5950 & 0.5158 & 0.5158 & 0.5175 \\
& Sortino ratio      & 0.9508 & 0.8180 & 0.8179 & 0.8226 \\
& Annualized returns  & 0.1217 & 0.1062 & 0.1062 & 0.1080 \\
& Annualized volatility & 0.1877 & 0.1866 & 0.1865 & 0.1893 \\
& Downside volatility  & 0.1175 & 0.1176 & 0.1176 & 0.1192 \\[1ex]
\multirow{5}{*}{\textbf{Communication services}} & Sharpe ratio       & 0.8342 & 0.6282 & 0.6327 & 0.6087 \\
& Sortino ratio      & 1.2913 & 0.9677 & 0.9745 & 0.9459 \\
& Annualized returns  & 0.2282 & 0.1546 & 0.1558 & 0.1530 \\
& Annualized volatility & 0.2303 & 0.2616 & 0.2304 & 0.2351 \\
& Downside volatility  & 0.1495 & 0.1690 & 0.1496 & 0.1512 \\[1ex]
\multirow{5}{*}{\textbf{Industrials}} & Sharpe Ratio       & 0.6944 & 0.6167 & 0.3991 & 0.3991 \\
& Sortino Ratio      & 1.3096 & 1.0976 & 0.7509 & 0.7509 \\
& Annualized Return  & 0.1517 & 0.1279 & 0.0818 & 0.0818 \\
& Annualized Volatility & 0.2043 & 0.1905 & 0.1808 & 0.1807 \\
& Downside Volatility  & 0.1186 & 0.1129 & 0.1067 & 0.1067 \\[1ex]
\multirow{5}{*}{\textbf{Consumer staples}} & Sharpe ratio       & 0.9601 & 0.7998 & 0.7996 & 0.7425 \\
& Sortino ratio      & 1.7167 & 1.5092 & 1.5088 & 1.3942 \\
& Annualized returns  & 0.1607 & 0.1385 & 0.1385 & 0.1313 \\
& Annualized volatility & 0.1610 & 0.1644 & 0.1644 & 0.1653 \\
& Downside volatility  & 0.0971 & 0.0981 & 0.0981 & 0.1000 \\[1ex]
\multirow{5}{*}{\textbf{Energy}} & Sharpe ratio       & 0.9784 & 0.6887 & 0.6886 & 0.8063 \\
& Sortino ratio      & 1.7598 & 1.2592 & 1.2591 & 1.5169 \\
& Annualized returns  & 0.2981 & 0.2065 & 0.2064 & 0.2282 \\
& Annualized volatility & 0.3122 & 0.2856 & 0.2856 & 0.2832 \\
& Downside volatility  & 0.1837 & 0.1704 & 0.1704 & 0.1681 \\[1ex]
\multirow{5}{*}{\textbf{Utilities}} & Sharpe ratio       & 0.3157 & 0.1953 & 0.1949 & 0.4153 \\
& Sortino ratio      & 0.6252 & 0.3978 & 0.3971 & 0.8049 \\
& Annualized returns  & 0.0874 & 0.0687 & 0.0686 & 0.1054 \\
& Annualized volatility & 0.2147 & 0.2191 & 0.2191 & 0.2171 \\
& Downside volatility  & 0.1266 & 0.1310 & 0.1310 & 0.1281 \\[1ex]
\multirow{5}{*}{\textbf{Materials}} & Sharpe ratio       & 0.8156 & 0.6544 & 0.6541 & 0.7749 \\
& Sortino ratio      & 1.5464 & 1.2818 & 1.2814 & 1.5298 \\
& Annualized returns  & 0.2247 & 0.1829 & 0.1829 & 0.2003 \\
& Annualized volatility & 0.2373 & 0.2119 & 0.2119 & 0.2200 \\
& Downside volatility  & 0.1356 & 0.1205 & 0.1205 & 0.1256 \\
\bottomrule
\end{tabularx}
}
\end{table}
\normalsize

\setcounter{table}{3}
\begin{table}[ht!]
{\fontsize{9.5}{10}\selectfont
\centering
\topcaption{\textbf{{Average metric across different portfolios constructed using LR, volatility, diversification ratio based on SD, and diversification based on VaR, continued.}\label{tab:results2}}}
\begin{tabularx}{\linewidth}{XX*{4}{d{3.6}}} 
\toprule
\textbf{Portfolio} & \textbf{Metric} & \multicolumn{1}{c}{\textbf{LR}} & \multicolumn{1}{c}{\textbf{Volatility}} & \multicolumn{1}{c}{$\text{\textbf{DR}}_{\text{\textbf{SD}}}$} & \multicolumn{1}{c}{$\text{\textbf{DR}}_{\text{\textbf{VaR}}}$} \\
\cmidrule(lr){1-1} \cmidrule(lr){2-2} \cmidrule(lr){3-3} \cmidrule(lr){4-4} \cmidrule(lr){5-5} \cmidrule(lr){6-6} 
\multirow{5}{*}{\textbf{Random mixed portfolio 1}} & Sharpe ratio       & 2.7954 & 1.6615 & 1.6618 & 1.8384 \\
& Sortino ratio      & 5.0747 & 3.4220 & 3.4229 & 3.8103 \\
& Annualized returns  & 0.3768 & 0.1864 & 0.1864 & 0.1927 \\
& Annualized volatility & 0.1256 & 0.0972 & 0.0972 & 0.0924 \\
& Downside volatility  & 0.0692 & 0.0485 & 0.0485 & 0.0462 \\[1ex]
\multirow{5}{*}{\textbf{Random mixed portfolio 2}} & Sharpe ratio       & 1.1692 & 0.3294 & 0.3297 & 0.3371 \\
& Sortino ratio      & 2.0943 & 0.6796 & 0.6800 & 0.7090 \\
& Annualized returns  & 0.2492 & 0.1074 & 0.1075 & 0.1056 \\
& Annualized volatility & 0.2006 & 0.1705 & 0.1705 & 0.1784 \\
& Downside volatility  & 0.1163 & 0.1009 & 0.1009 & 0.1055 \\[1ex]
\multirow{5}{*}{\textbf{Random mixed portfolio 3}} & Sharpe ratio       & 1.2149 & 0.8924 & 0.8925 & 0.8150 \\
& Sortino ratio      & 2.2863 & 1.7493 & 1.7496 & 1.5559 \\
& Annualized returns  & 0.2408 & 0.1508 & 0.1508 & 0.1564 \\
& Annualized volatility & 0.2062 & 0.1766 & 0.1766 & 0.1875 \\
& Downside volatility  & 0.1190 & 0.1038 & 0.1038 & 0.1094 \\[1ex]
\multirow{5}{*}{\textbf{Random mixed portfolio 4}} & Sharpe ratio       & 1.0826 & 1.0450 & 1.0446 & 0.8131 \\
& Sortino ratio      & 2.0263 & 1.9618 & 1.9612 & 1.5295 \\
& Annualized returns  & 0.2196 & 0.1776 & 0.1775 & 0.1577 \\
& Annualized volatility & 0.2036 & 0.1794 & 0.1794 & 0.1957 \\
& Downside volatility  & 0.1152 & 0.1036 & 0.1036 & 0.1122 \\[1ex]
\multirow{5}{*}{\textbf{Random mixed portfolio 5}} & Sharpe ratio       & 0.9731 & 0.2399 & 0.2401 & 0.3330 \\
& Sortino ratio      & 1.8215 & 0.5130 & 0.5136 & 0.6348 \\
& Annualized returns  & 0.2046 & 0.0680 & 0.0680 & 0.0869 \\
& Annualized volatility & 0.2035 & 0.1815 & 0.1815 & 0.1888 \\
& Downside volatility  & 0.1173 & 0.1082 & 0.1082 & 0.1132 \\[1ex]

\multirow{5}{*}{\textbf{High volatility}} & Sharpe ratio       & 0.5100 & 0.4392 & 0.4402 & 0.4768 \\
& Sortino ratio      & 0.8342 & 0.7164 & 0.7177 & 0.7833 \\
& Annualized returns  & 0.1144 & 0.0883 & 0.0886 & 0.1026 \\
& Annualized volatility & 0.1923 & 0.2243 & 0.1923 & 0.2032 \\
& Downside volatility  & 0.1221 & 0.1416 & 0.1221 & 0.1277 \\[1ex]
\multirow{5}{*}{\textbf{Low volatility}} & Sharpe ratio       & 0.7421 & 0.4510 & 0.4511 & 0.6631 \\
& Sortino ratio      & 1.3583 & 0.8512 & 0.8514 & 1.2525 \\
& Annualized returns  & 0.1474 & 0.1083 & 0.1084 & 0.1447 \\
& Annualized volatility & 0.1765 & 0.1730 & 0.1730 & 0.1775 \\
& Downside volatility  & 0.1026 & 0.1017 & 0.1017 & 0.1016 \\[1ex]
\multirow{5}{*}{\textbf{Mixed volatility}} & Sharpe ratio       & 1.5571 & 0.6973 & 0.6972 & 0.9459 \\
& Sortino ratio      & 3.1229 & 1.3418 & 1.3415 & 1.7901 \\
& Annualized returns  & 0.3167 & 0.1387 & 0.1387 & 0.1903 \\
& Annualized volatility & 0.2279 & 0.1637 & 0.1637 & 0.1749 \\
& Downside volatility  & 0.1311 & 0.0983 & 0.0983 & 0.1034 \\
\bottomrule
\end{tabularx}
}
\end{table}
\normalsize

\section{Discussion\label{sec:discussion}}

The methodology used in this study effectively demonstrated the potential of LR as a novel diversification metric. We examined how LR relates to established measures like Markowitz's volatility and the diversification ratios based on SD and VaR. While statistically related, LR and these traditional metrics are not identical, which underscores LR's ability to provide additional insights by capturing diversification dimensions beyond simple return correlations. This distinction is crucial, as lexical ratio's use of textual data enables it to identify hidden relationships and correlations between assets that conventional measures might miss.

Traditional metrics often fail to account for underlying complexities in asset relationships, especially during market stress or volatility periods. In contrast, LR does not depend on restrictive assumptions about the data and leverages a broader set of information, which allows it to offer a more comprehensive view of portfolio diversification. Its desirable mathematical properties, such as scale invariance and maximality, ensure that LR remains robust across a variety of portfolio configurations, capturing insights that return-based measures often overlook. The numerical results of our robustness analysis confirmed this.

The real-world testing of LR, particularly the out-of-sample analysis on S\&P 500 portfolios, provided much-needed validation of its practical application. By applying a rolling-window approach, we tested the performance of LR across multiple market conditions, demonstrating that it consistently outperformed traditional metrics in identifying well-diversified portfolios, even in sectors prone to volatility or shocks.

Although our dataset, spanning from 2018 to 2024, had some limitations regarding linguistic data availability, this did not significantly detract from the study's findings. With the increasing abundance of financial data, the potential for LR to perform even better is clear. 

As introduced, LR has demonstrated promise in assessing portfolio diversification by leveraging textual information from financial news. However, several potential extensions could enhance its application, particularly by incorporating time decay and risk-adjusted approaches and using them to develop systematic risk metrics. These extensions would make LR more adaptable to the dynamic nature of financial markets and better reflect the risk factors that influence portfolio performance.

One potential enhancement is incorporating time decay to adjust the influence of older news items. As time passes, financial news loses relevance, and its impact on portfolio decisions diminishes. By applying a decay factor, the lexical ratio can more effectively discount older information, giving greater weight to recent news. The decay weight for each headline can be formulated as
\[
e^{-\lambda \cdot t_i}
\]
where \( \lambda \) is the decay rate that controls how quickly older news loses significance, and \( t_i \) denotes the time difference between the current time and the publication date of the $i^{th}$ headline. Incorporating this decay into the LR calculation results in a \textit{time-sensitive lexical ratio}, which better captures the shifting relevance of news in response to changing market conditions. This extension ensures that more recent developments have a greater impact on the diversification assessment, making LR more responsive to new information.

Another promising extension involves adjusting LR to account for risk by assigning different weights to specific words or phrases indicative of financial risk. For instance, terms such as ``bankruptcy,'' ``recession,'' or ``downgrade'' may significantly affect portfolio stability. By giving more weight to these risk-related terms, the lexical ratio can be adjusted to better capture the potential downside risks inherent in a portfolio. The weight assigned to each term \( j \) can be
\[
\begin{cases}
    \gamma & \text{if term } j \text{ is a risk keyword} \\
    1 & \text{otherwise}
\end{cases} ,
\]
where \( \gamma > 1 \) is a \textit{boost factor} used to amplify the influence of terms associated with financial risk, and risk keywords are identified through sentiment analysis or by selecting terms that are known indicators of negative market events. The resulting risk-sensitive LR can be computed by incorporating these adjusted term weights into the Shannon entropy calculation:
\[
\text{LR}_{\text{risk}} = \frac{-\sum_j p_j \log(p_j)}{\log(m)}
\]
where \( p_j \) represents the adjusted probability of term \( j \), and \( m \) is the total number of terms in the vocabulary. By weighting risk-related terms more heavily, the risk-sensitive LR provides a risk-aware diversification metric that is more sensitive to potentially negative events.

A further extension involves developing systematic risk metrics using LR to assess how diversification reduces exposure to systematic risk across sectors. Cross-sector diversification is a key strategy for mitigating systematic risk, as high correlations among assets within a sector may increase overall portfolio risk. By applying LR across different sectors, we can measure the impact of news that affects multiple sectors and examine cross-sector correlation patterns over time. High correlation across LR values for sectors indicates increased systematic risk due to common macroeconomic factors. Monitoring LR in this context can provide valuable insights into detecting systematic risk.


Future research could involve extending the optimization framework to include these parameters, potentially employing multi-objective optimization to balance maximizing LR while minimizing risk exposure. This would reflect the real-world trade-offs portfolio managers face in balancing potential returns with acceptable levels of risk.

\section{Concluding remarks\label{sec:conclusion}}

The lexical ratio introduces a fresh perspective to portfolio diversification metrics, fundamentally changing how we assess risk and diversification. Unlike traditional metrics that rely solely on asset returns, LR integrates additional available information from textual data, offering a more complete view of portfolio composition. This approach allows us to avoid making restrictive assumptions about the underlying data, making LR especially effective in capturing relationships between assets that traditional correlation-based methods often miss.

Our study demonstrated that LR offers valuable insights into portfolio composition and hidden correlations between assets that conventional measures, such as Markowitz's volatility and diversification ratios, fail to capture. Through extensive real-world testing on S\&P 500 portfolios, we illustrated that LR is not only related meaningfully to established metrics but also provides unique diversification benefits, particularly in volatile market conditions or when faced with sector-specific shocks.

Furthermore, the robustness of the LR metric across various portfolio structures, whether sector-specific, random, or volatility-based, underscored its versatility. While equal-weight strategies may perform competitively in certain cases, LR's ability to integrate textual information offers a more sophisticated approach for portfolio managers seeking to balance risk and optimize returns.

This article's findings pave the way for further exploration of NLP techniques in financial decision-making, offering a framework that can be extended to more complex asset classes and market environments. As financial markets become increasingly data-driven, the lexical ratio presents a forward-thinking, information-rich approach to portfolio diversification.

\section*{Acknowledgements}

\Acknowledgements

\section*{Data statement}
This study utilizes news headline data from the \texttt{eodhd} package. Stock return data is sourced from the \texttt{yfinance} library. All code and data used in the production of this paper can be accessed via \href{https://github.com/frazmohs/Lexical_Ratio_Preprint}{GitHub}.

\section*{Declaration of competing interests}

The authors declare they have no known competing financial interests or personal relationships that could have appeared to influence the work reported in this article.

\setlength{\bibsep}{6pt}
\titlespacing*{\section}{0pt}{6pt}{12pt}
\singlespacing
\begin{spacing}{0.0}
\bibliographystyle{apa-good}
\bibliography{references}
\end{spacing}
\vspace{0.3cm}

\doublespacing
\appendix

\section{Proof of scale invariance of the lexical ratio\label{app:proof}}

To prove that LR is scale-invariant, consider multiplying all asset weights \(w_i\) by a positive constant factor \(c > 0\):
\[
w_i' = c \, w_i
\]
Substituting the new weights into the expression for LR yields a new version of the lexical ratio:
\[
\text{LR}' = - \frac{1}{\log(m)} \sum_{k=1}^{m} \left( \frac{\sum_{i=1}^{n} (c \, w_i) \cdot c_{ik}}{\sum_{i=1}^{n} \sum_{j=1}^{m} (c \, w_i) \, c_{ij}} \right) \log \left( \frac{\sum_{i=1}^{n} (c \, w_i) \, c_{ik}}{\sum_{i=1}^{n} \sum_{j=1}^{m} (c \, w_i) \, c_{ij}} \right)
\]
Since \(c > 0\), the constant \(c\) cancels out in both the numerator and denominator:
\[
\text{LR}' = - \frac{1}{\log(m)} \sum_{k=1}^{m} \sum_{k=1}^{m} \left( \frac{\sum_{i=1}^{n} w_i \, c_{ik}}{\sum_{i=1}^{n} \sum_{j=1}^{m} w_i \, c_{ij}} \right) \log \left( \frac{\sum_{i=1}^{n} w_i \, c_{ik}}{\sum_{i=1}^{n} \sum_{j=1}^{m} w_i \, c_{ij}} \right),
\]
leading to 
\[
\text{LR}' = \text{LR}.
\]
This proves that the lexical ratio is scale invariant.

\end{document}

%% file: info.tex
\newcommand{\Title}{The lexical ratio: \\ A new perspective on portfolio diversification}

\newcommand{\Acknowledgements}{B\'egin wishes to acknowledge the financial support of the Natural Sciences and Engineering Research Council of Canada and Simon Fraser University.}

\newcommand{\Authors}{Sayyed Faraz Mohseni\footnote{Independent researcher without institutional affiliation, e-mail: farazmohseni15@gmail.com.} \qquad Hamid Arian\footnote{School of Administrative Studies, York University, 96 The Pond Road, North York, Ontario, Canada, M3J 2S5, e-mail: harian@yorku.ca.} \qquad Jean-Fran\c cois B\'egin\footnote{Department of Statistics and Actuarial Science, Simon Fraser University, 8888 University Drive, Burnaby, British Columbia, Canada, V5A 1S6, e-mail: jbegin@sfu.ca.} } 

\newcommand{\Abstract}{
Portfolio diversification, traditionally measured through asset correlations and volatility-based metrics, is fundamental to managing financial risk. However, existing diversification metrics often overlook non-numerical relationships between assets that can impact portfolio stability, particularly during market stresses. This paper introduces the lexical ratio (LR), a novel metric that leverages textual data to capture diversification dimensions absent in standard approaches. By treating each asset as a unique document composed of sector-specific and financial keywords, the LR evaluates portfolio diversification by distributing these terms across assets, incorporating entropy-based insights from information theory. We thoroughly analyze LR's properties, including scale invariance, concavity, and maximality, demonstrating its theoretical robustness and ability to enhance risk-adjusted portfolio returns. Using empirical tests on S\&P 500 portfolios, we compare LR's performance to established metrics such as Markowitz's volatility-based measures and diversification ratios. Our tests reveal LR's superiority in optimizing portfolio returns, especially under varied market conditions. Our findings show that LR aligns with conventional metrics and captures unique diversification aspects, suggesting it is a viable tool for portfolio managers. 
}

\newcommand{\Keywords}{Diversification; Shannon entropy; Portfolio; Dependence; Natural language processing.}

\newcommand{\JEL}{G11, G14.}